\documentclass{article}

\usepackage{arxiv}

\usepackage[utf8]{inputenc} 
\usepackage[T1]{fontenc}    
\usepackage{hyperref}       
\usepackage{url}            
\usepackage{booktabs}       
\usepackage{amsfonts}       
\usepackage{nicefrac}       
\usepackage{microtype}      
\usepackage{lipsum}
\usepackage{graphicx}

\title{Dispersion forces stabilise ice coatings at certain gas hydrate interfaces which prevent water wetting}

\author{M.~Bostr{\"o}m\thanks{Centre for Materials Science and Nanotechnology, Department of Physics, University of Oslo, P. O. Box 1048 Blindern, NO-0316 Oslo, Norway}\\
Department of Energy and Process Engineering\\Norwegian University of Science and Technology\\NO-7491 Trondheim, Norway\\\texttt{mathias.a.bostrom@ntnu.no}
 \And
R.~Corkery\thanks{Applied Physical Chemistry, KTH Royal Institute of Technology, SE 100 44 Stockholm, Sweden}\\
Surface and Corrosion Science\\Department of Chemistry\\KTH Royal Institute of Technology\\SE 100 44 Stockholm, Sweden\\\texttt{corkery@kth.se}
\And
E.~R.~A.~Lima\\Programa de P{\'{o}}s-gradua{\c{c}}{\~{a}}o em Engenharia Qu{\'{i}}mica\\Universidade do Estado do Rio de Janeiro\\CEP 20550-013, Rio de Janeiro RJ, Brazil
\And
O.~I.~Malyi\\Centre for Materials Science and Nanotechnology\\Department of Physics\\University of Oslo\\P. O. Box 1048 Blindern\\NO-0316 Oslo, Norway
\And
S.~Y.~Buhmann~\thanks{Freiburg Institute for Advanced Studies, Albert-Ludwigs-Universit{\"a}t Freiburg, Albertstr. 19, 79104 Freiburg, Germany}\\
Physikalisches Institut\\Albert-Ludwigs-Universit{\"a}t Freiburg\\Hermann-Herder-Str. 3\\79104 Freiburg, Germany
\And
C.~Persson\\Centre for Materials Science and Nanotechnology\\Department of Physics\\University of Oslo\\P. O. Box 1048 Blindern\\NO-0316 Oslo, Norway
\And
I.~Brevik\\
Department of Energy and Process Engineering\\Norwegian University of Science and Technology\\NO-7491 Trondheim, Norway
\And
D.~F.~Parsons\\School of Engineering and IT\\Murdoch University\\90 South St\\Murdoch, WA 6150, Australia\\
\texttt{d.parsons@murdoch.edu.au}
\And
J.~Fiedler~\thanks{Centre for Materials Science and Nanotechnology, Department of Physics, University of Oslo, P. O. Box 1048 Blindern, NO-0316 Oslo, Norway}\\
Physikalisches Institut\\Albert-Ludwigs-Universit{\"a}t Freiburg\\Hermann-Herder-Str. 3\\79104 Freiburg, Germany\\
\texttt{johannes.fiedler@physik.uni-freiburg.de}}

\begin{document}
\maketitle

\begin{abstract}
Gas hydrates formed in oceans and permafrost occur in vast quantities on Earth representing both a massive potential fuel source and a large threat in climate forecasts. They have been predicted to be important on other bodies in our solar systems such as Enceladus, a moon of Saturn. CO$_2$-hydrates likely drive the massive gas-rich water plumes seen and sampled by the spacecraft Cassini, and the source of these hydrates is thought to be due to buoyant gas hydrate particles. Dispersion forces cause gas hydrates to be coated in a 3-4\,nm thick film of ice, or to contact water directly, depending on which gas they contain. These films are shown to significantly alter the properties of the gas hydrate clusters, for example, whether they float or sink. It is also expected to influence gas hydrate growth and gas leakage.
\end{abstract}


\section{Introduction}

 Gas hydrates are systems consisting of water and gas molecules 
forming 
a solid ice structure. Such systems can naturally be found in  ice-cold water~\cite{doi:10.1029/94GL01858};  in particular, they can occur in permafrost~\cite{NaturalGas2003}, sediments~\cite{doi:10.1029/2007GC001920},  and below the oceans in the seabed~\cite{doi:10.1002/2016GL068656}. For the latter, there are particularly interesting examples where gas hydrates are considered important in connection with 
planetary processes and the implications for life. The  aqueous ocean-bearing moons Europa and Enceladus are perhaps the best examples in our solar system beyond Earth where gas hydrates are formed in salty oceans that are favourable for life~\cite{Nimmo2016}. On Mars, methane distribution is associated with subterranean water, implying the presence of methane hydrates~\cite{Fonti2010}. 
On Enceladus giant plumes of erupted gases are observed and the composition directly measured to be water, salts and volatile gases including CO$_2$, CO, N$_2$, H$_2$S and methane~\cite{Waite2006,Postberg2011}. Several hypotheses consider gas hydrates to be important for the creation of volatile enriched plumes and for the composition of ice layers beneath and/or entrained into, or sprayed onto the outer surface of Enceladus~\cite{Bouquet2015,Matson2012,Matson2018}.
In particular, type II gas hydrates on Enceladus and Europa are calculated to be less dense than water and can float in their respective oceans. They are thereby available for incorporation into the overlying thick ice layer of each icy moon. Type I CO$_2$ hydrates are at a density where their positive or negative buoyancy is uncertain~\cite{McKay2003,PrietoBallesteros2005,Mousis2013,Safi2017,Bouquet2015}. However, if a layer of water ice forms on these gas hydrates in the presence of ice cold liquid water, then the growth of such hydrate crystals may be limited by the capping effect. This may have an impact on their buoyancy, and thus on the hypothesized composition of the ice layers in Enceladus and Europa, with obvious implications for the composition of their plumes and their potential to sample the underlying oceans and any harboured life.
\begin{figure}[b]
    \centering
    \includegraphics[width=0.5\textwidth]{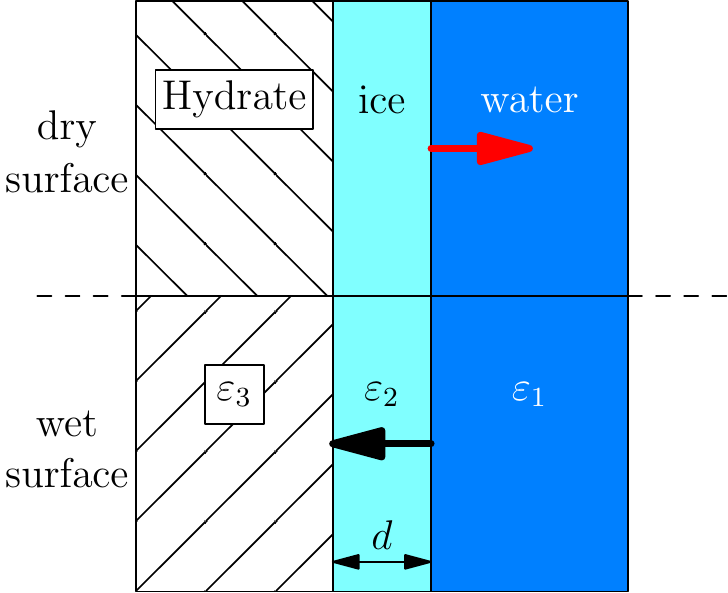}
    \caption{(Color online). Schematic figure of the considered arrangement. A gas hydrate surface ($\varepsilon_3$) on the left, separated by an ice layer ($\varepsilon_2$) of thickness $d$ from a water layer ($\varepsilon_1$). A dry surface feels a repulsive Casimir force at the ice-water interface which yields a stable ice interface. In contrast, an attractive force results in a wet surface due to the vanishing of the ice interface.}
    \label{fig:scheme}
\end{figure} 

On  Earth,  methane hydrates occur naturally and in engineered situations. Large reservoirs of methane hydrates occur in sediments of deep oceans basins, at shallower depths in the sediments of arctic sea shelves, and in deep permafrost regions. In all these cases, the understanding of whether  a layer of  ice forms on the hydrate has implications for exploration and production of fossil fuels, and also for understanding the potential for methane contribution  to greenhouse gases as the planet becomes warmer.

In all of the  contexts above,
hydrates are usually surrounded by ice cold water. Depending on the gas hydrate structure with respect to contributions and volume fractions, it turns out that some hydrates form an ice interface to water, whereas others do not. The latter have a wet surface.  The ones with a gas hydrate-ice interface may be considered to have a dry surface. The prediction of the wet or dry surface cannot be made easily. In the present paper we address this issue by considering a planar three-layer system, as depicted in Fig.~\ref{fig:scheme}, namely   a gas hydrate layer, an ice sheet, and a water layer. Thus, we assume that initially all hydrates are covered by an ice interface. We estimate the Casimir force acting on the outside at the ice-water interface, i.e.,  the  pressure acting on the system. Depending on the sign of the Casimir force,  it will work towards growth or melting of the thin ice sheet.  We assume that the temperature is at the triple point of water.  An attractive pressure acting on the ice layer  thus results in a melting of the ice sheet~\cite{Landau,Nilsson2015,doi:10.1063/1.1733772}.  It will  simply vanish.
This kind of consideration is not new. In the past, ice melting  at the triple point with a nano-sized  film of water was discussed~\cite{Elbaum}. It was found that a  thin   water film is  energetically favourable up to a certain thickness where it has an energy minimum~\cite{Elbaum,Elbaum1993,Wilen,Dash1995,Wettlaufer}. The inclusion of retardation resulted in incomplete melting while a non-retarded approximation predicted complete melting for an ice surface at the triple point of water~\cite{Elbaum}.

 Here we apply Lifshitz theory to estimate the energy of the hydrate-ice-system as a function of ice thickness, and show that for some gas hydrates the ice film is stabilised at a thickness of 3-4 nm, while for other gas hydrates the ice film is unstable, resulting in direct wetting.

\section{Materials and Methods}

\subsection{Dispersion forces between solid bodies}

 The Casimir interaction energy $F(d)$ (also known as Lifshitz free energy) per unit area between material 1 (water) with dielectric function $\varepsilon_1$ and material 3 (gas hydrate), $\varepsilon_3$, separated by the distance $d$ across medium 2 (ice), $\varepsilon_2$ as depicted in Fig.~\ref{fig:scheme} can be written at temperature $T$   as~\cite{Dzya,Pars,Ninhb,Buhmann12a}  
\begin{equation}
F(d) = \sum\limits_{n = 0}^\infty  {}' \left[g^{\rm TM}\left( \xi_n \right) + g^{\rm TE}\left( \xi_n \right)\right]\,,
\label{equ5}
\end{equation}
where $g^{\rm TX}(\xi)$ (TX=TM, TE) denotes the trace over the scattering for transverse magnetic (TM) and transverse electric (TE) Green's function. This fundamental solution comes from  the vector Helmholtz equation for the electric field. The primed sum denotes that the $n=0$ term  is weighted by a factor one half.  At  finite temperature these functions are evaluated at   the discrete Matsubara frequencies $\xi_n = 2\pi n k_{\rm B} T/\hbar$~\cite{Buhmann12b}. The systems in this study, as mentioned,  are all studied at the triple point of water. For the considered three layer system, the traces over the scattering Green's functions, including  multiple reflection in the center layer, can be written  (in cgs units) as
\begin{equation}
g^{\rm TX}\left( \xi_n \right) = \frac{1}{\beta }\int \frac{\mathrm d^2q}{\left( 2\pi  \right)^2}\ln \left\{ 1 - \mathrm e^{ - 2\gamma_2d} r_{12}^{\rm TX} r_{32}^{\rm TX}\right\}\,,
\label{equ_6}
\end{equation}
with $\beta = 1/(k_{\rm B}T)$, and the Fresnel reflection coefficients are
\begin{equation}
r_{i2}^{\rm TM} =   \frac{\varepsilon_i\gamma_2 -\varepsilon_2\gamma_i}{\varepsilon_i\gamma_2 +\varepsilon_2\gamma_i}  \,,
\label{Equ7}
\end{equation}
for TM waves and
\begin{equation}
    r_{i2}^{\rm TE} = \frac{\gamma_2-\gamma_i}{\gamma_2+\gamma_i} \,,
    \label{Equ8}
\end{equation}
for TE waves. We have introduced the imaginary part of the transverse wave vector $\gamma_i^2 = q^2+\xi^2 \varepsilon_i/c^2$.   We assume nonmagnetic media.

\subsection{Material Modelling}

\begin{figure}\centering
\includegraphics[width=0.5\columnwidth]{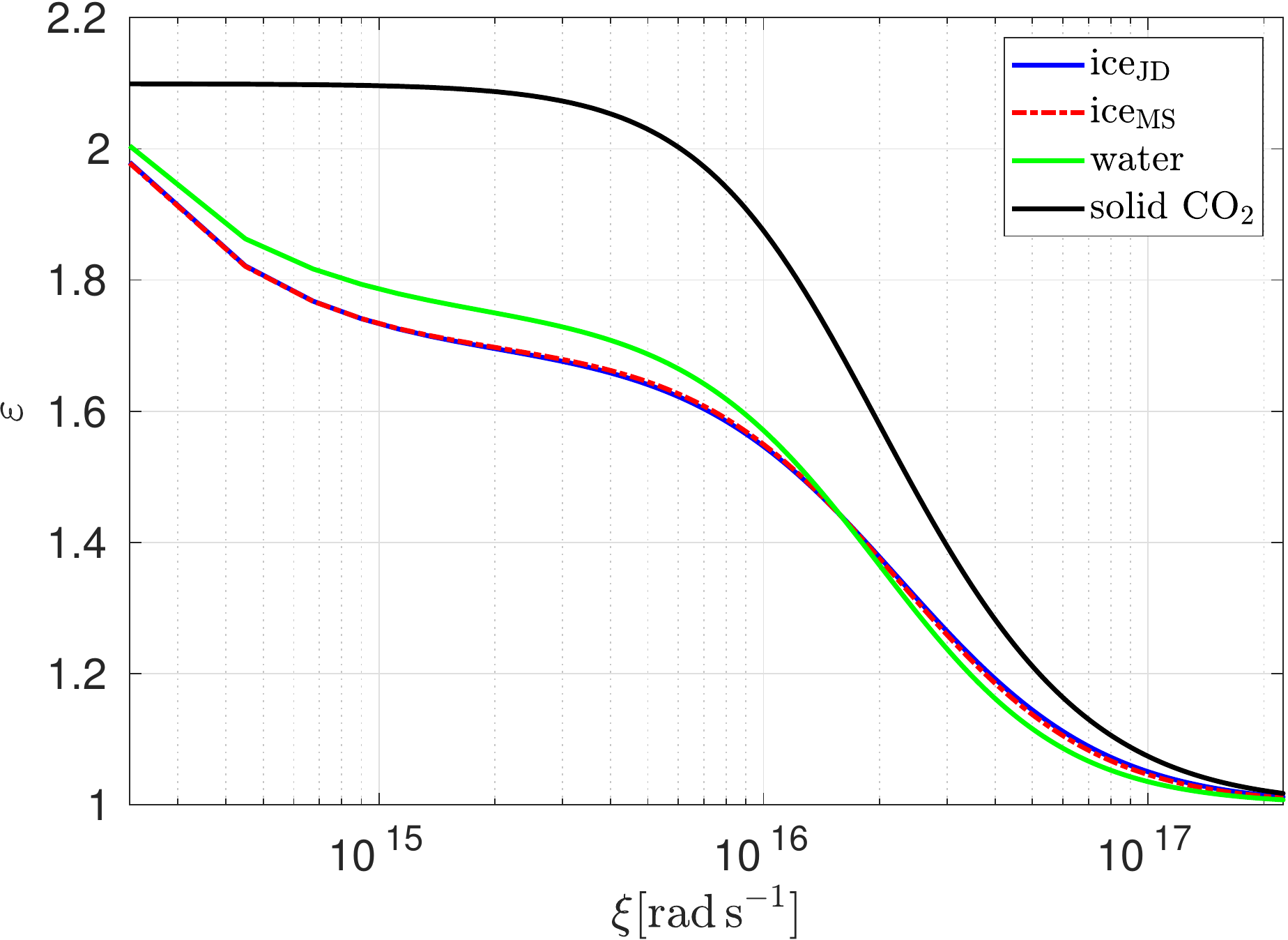}
\caption{(Color online) The dielectric functions of crystalline CO$_2$\cite{Thiyam2016}, both ice models and water~\cite{Elbaum} at 273.16 K.} 
\label{diel_CrystallineCO2}
\end{figure}

Dielectric functions (at imaginary frequencies)  were taken from Elbaum and Schick \cite{Elbaum} using the data from Daniels \cite{Ref12} and labeled by ice$_\mathrm{JD}$ and from Seki et al.\cite{Ref13} (ice$_\mathrm{SM}$) for ice ($\varepsilon_2$)
and from Elbaum and Schick\cite{Elbaum} for water ($\varepsilon_1$).
 These dielectric functions  are for a system at the triple point of water, close to zero degrees Celsius at low pressure. The final results for ice melting~\cite{Elbaum,Elbaum1993,Wilen,Dash1995,Wettlaufer,Thiyam2016,Bostr2016} 
  and water freezing\,\cite{Elbaum2,MBPhysRevB02017} are sensitive to the dielectric functions of ice and water since these are extremely similar when the water is in equilibrium with the ice. We show in Fig.~\ref{diel_CrystallineCO2} the dielectric functions for crystalline CO$_2$, water and ice.

A model for the dielectric function of  a gas  hydrate ($\varepsilon_3$)  is derived using the Lorentz-Lorenz model~\cite{Aspnes} with the mixing scheme specifically for gas hydrates taken from Bonnefoy et al.\cite{BONNEFOY2005336,Bonnefoy}
\begin{equation}
\varepsilon_3=\frac {1+2 \Gamma}{1- \Gamma} \, , \label{equ1} 
\end{equation}
with 
\begin{equation}
 \Gamma={\frac {\varepsilon_{2}-1} {\varepsilon_{2}+2}} \left({\frac{n_{\rm wh}}{n_{\rm i} }}\right)+{\frac{4 \pi \alpha_{\rm M} n_{\rm M}} {3}}\, ,
\label{equ2}
\end{equation}
which means that the dominating factors for the dielectric function of gas hydrates are the ice polarisability weighted by density of water in the hydrate relative to pure ice, and the polarisabilities of different gas molecules weighted by their corresponding densities. The mass density of water in pure ice is 0.9167 g/cm$^{3}$~\cite{CRChandbook}, giving the number  density of water molecules in pure ice as $n_{\rm i}= 3.06\times 10^{-2}$ \,{\AA}$^{-3}$. The number densities of gas molecules ($n_{\rm M}$)  and water molecules 
($n_{\rm wh}$)   
in different gas hydrate structures are tabled in Tab.~\ref{EduTable} with the water/gas number density ratio
\begin{equation}
    p= \frac{n_{\rm wh}}{n_{\rm M}} \, .
\end{equation}

\begin{table*}[b]
\centering
\caption{Hydrate mass densities ($\rho_{\rm h}$) and number densities of water ($n_{\rm wh}$) and gas molecules ($n_{\rm M}$) in different gas hydrates. The water/gas number density ratio is denoted $p$.}

 \begin{tabular}{l|c|c|c|c}
    Gas molecule & $p$& $\rho_{\rm h}$ (g/cm$^{3}$)  & $n_{\rm M}$ (\AA$^{-3}$)   & $n_{\rm wh}$ (\AA$^{-3}$)  \\ \hline

CO$_2$,\,\cite{FerdowsOta}    &5.75& 1.13 &    $4.61\times10^{-3}$            & $2.65\times10^{-2}$                          \\
CO$_2$,\,\cite{FerdowsOta} &7.67    &1.05  &  $3.47\times10^{-3}$       & $2.66\times10^{-2}$               \\
CO$_2$,\,\cite{Makogon} &6.0  &1.117 & $4.42\times10^{-3}$     & $2.65\times10^{-2}$                \\
CH$_4$,\,\cite{NaturalGas2003}     &5.75 & 0.90     & $4.53\times10^{-3}$&        $2.60\times10^{-2}$                            \\
CH$_4$,\,\cite{Makogon}     &6.0 &0.91     &   $4.41\times10^{-3}$&       $2.65\times10^{-2}$             \\
  H$_2$S,\,\cite{Makogon}   &7.0  &1.044      &$3.92\times10^{-3}$      &  $2.75\times10^{-2}$        \\
N$_2$,\,\cite{Makogon}    & 6.0    &0.995   & $4.4\times10^{-3}$      &$2.64\times10^{-2}$ 
  \end{tabular}
\label{EduTable}
\end{table*}

Quantum chemical calculations of dynamic polarisabilities at discrete frequencies were represented at arbitrary imaginary frequencies $i\xi$ by fitting to the oscillator model,
\begin{equation}
\alpha_{\rm M}(i \xi)=\sum_j\frac{\alpha_j}{1+(\xi/\omega_j)^2}
\label{Equpol}
\end{equation}
A 5-mode fit has previously been found to describe the dynamic polarisability accurately to a 0.02\% relative error \cite{ParsonsNinham2010dynpol}. The adjusted parameters for a 5-mode model for CO$_2$, CH$_4$, N$_2$, and H$_2$S are given in our recent work~\cite{Johannes}. Quantum calculations on which the fits were based were taken at a coupled-cluster singles and doubles (CCSD) level of theory \cite{HampelPetersonWerner1992} using aug-cc-pVQZ basis sets \cite{WoonDunning1993}.

\subsection{Product of Reflection Coefficients}

\begin{figure}
\centering
 \includegraphics[width=0.5\columnwidth]{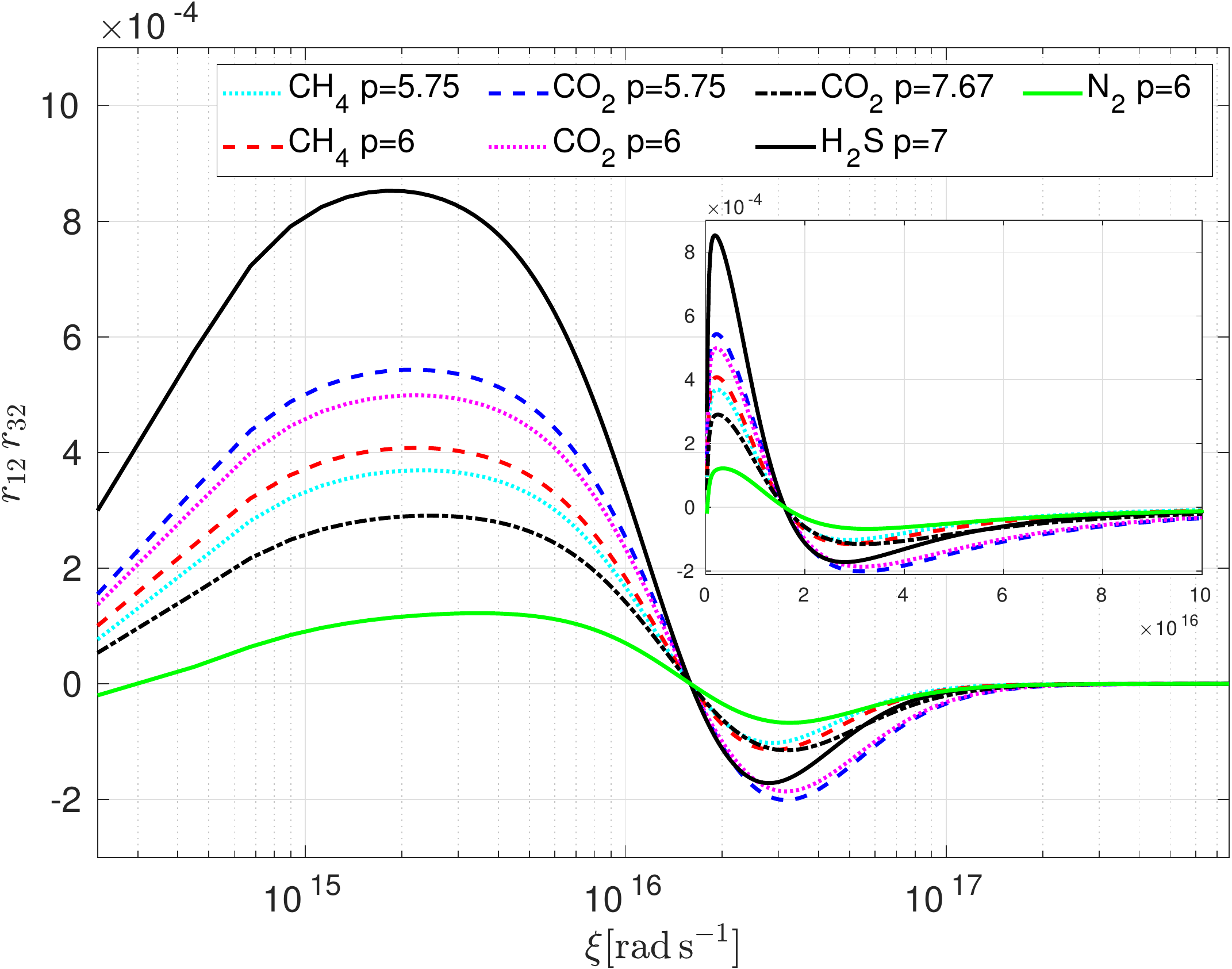}

 \caption{(Color online) Product of the non-retarded reflection coefficients for the two interfaces. Here we show these for the three CO$_2$ gas hydrates and the two CH$_4$ gas hydrates that we consider (the inset shows the same on a linear scale). }
 \label{reflections}
\end{figure}
As 
can be observed in Eq.~(\ref{equ_6}), the Casimir force is determined by the product of  reflection coefficients at both interfaces, 
summed over all frequencies. Thus, the magnitude and sign is given by the balance of 
areas enclosed by these curves above and below the frequency axis.
This behaviour is illustrated in Fig.~\ref{reflections}, where the products of the non-retarded Fresnel  coefficients (TM mode)  are shown each given by
\begin{equation}
r_{i2} =   \frac{\varepsilon_i -\varepsilon_2}{\varepsilon_i +\varepsilon_2}.
\label{Equ4}
\end{equation}

One can get insights from this quantity  also in cases where retardation matter. 
Negative values for the product shown in Fig.~\ref{reflections} (larger in magnitude for CO$_2$ hydrates than for CH$_4$ hydrates) for high frequencies contribute to repulsion.   The crossing point at $1.6\times 10^{16}\,\rm{ rad\,s}^{-1}$ where $r_{12}r_{32}=0$ corresponds to the frequency where the dielectric function of ice crosses that of pure water, seen in Figure~\ref{diel_CrystallineCO2}.
For nonretarded, small film thicknesses, the respective sum over all frequencies
(with many more terms for high frequencies than  for low frequencies) gives the net sign for the free energy of very thin ice films. 
Retardation favours the small-frequency contributions and hence
screens out high frequency (repulsive) contributions for thicker ice films. It turns out that already for film thicknesses as thin as a few nanometers retardation is important for ice-water related systems~\cite{Elbaum}. The net sign in our case is not trivial, and we will demonstrate later that CO$_2$ and N$_2$ hydrates in water behave differently from CH$_4$ and H$_2$S hydrates in water.

\section{Results}

\subsection{Gas hydrate specific ice formation}
While it is  well known that water can start to freeze from its surface when the temperature goes to zero degrees Celsius,  Elbaum and Schick~\cite{Elbaum2} predicted that dispersion forces do not play a role in this mechanism. In fact, they found that a thin ice film on the surface would have its energy minimum for zero ice film thickness which would not result in  surface freezing on open water surfaces. The underlying mechanism for why ice growth actually occurs at the surface is that large ice structures float with a certain fraction above a water surface due to the lower density of ice. In contrast to their results,  we have  found that buoyancy combined with dispersion and double layer forces establish an equilibrium where large ice particles float on the surface while small (micron-sized) ice particles  are  trapped at a distance below a water surface~\cite{Thiyam2018}.  Further, it was shown that ice formation can be induced by dispersion forces near silica-water interfaces (where silica can be used as a model for rock material)~\cite{MBPhysRevB02017}.

\begin{figure}
\centering
\includegraphics[width=0.5\columnwidth]{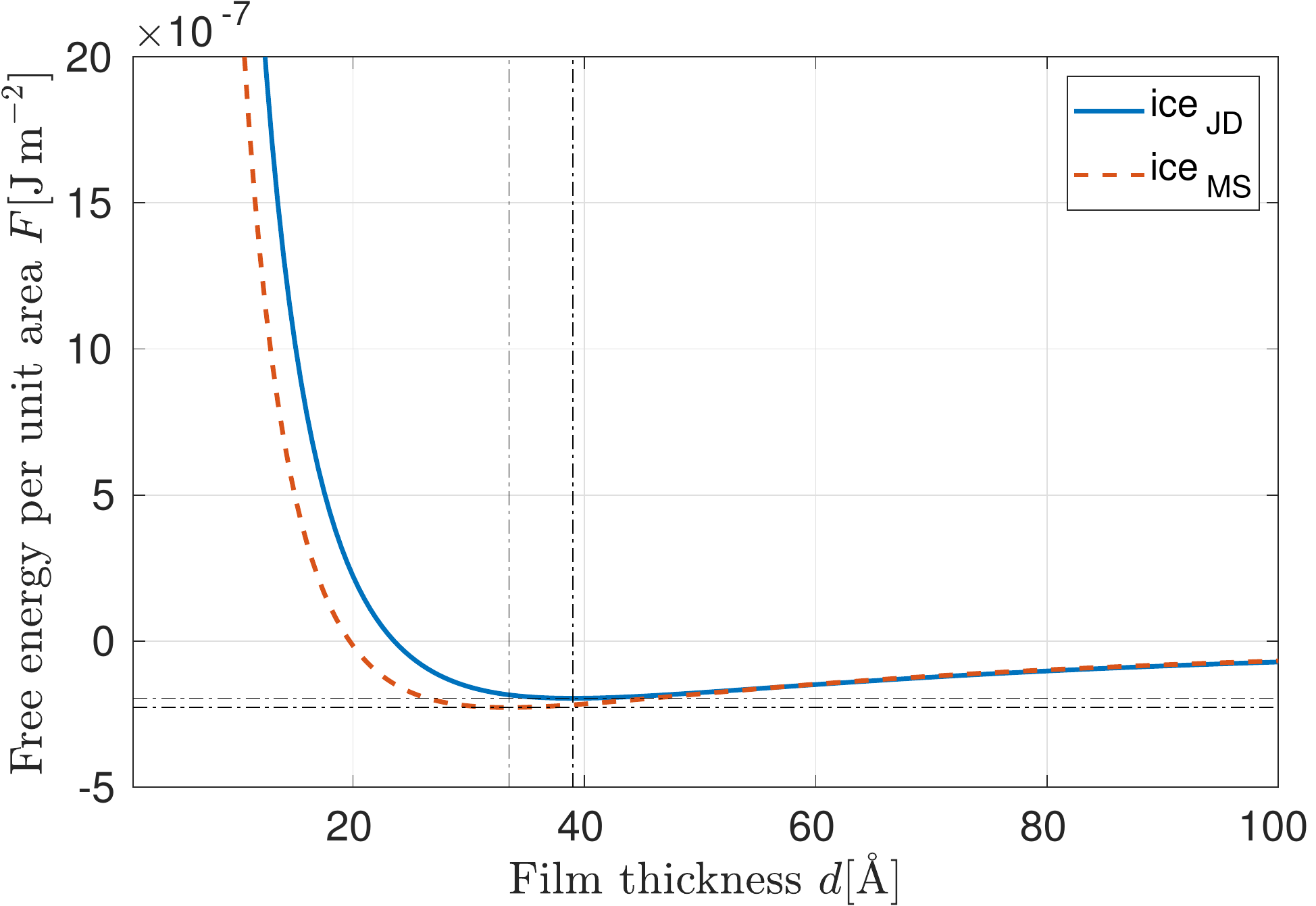}
\caption{(Color online) The free energy per unit area (at 273.16 K)  as a function of ice film thickness on the boundary between the surface of a crystalline CO$_2$ structure and  liquid water. It is predicted that at equilibrium an ice film around {$d$ = 33\,\AA} for the ice model from M. Seki \textit{et al.} and {$d$ = 39\,\AA} for J. Daniels' model. } 
\label{FreeziingCrystalCO2}
\end{figure}

Before presenting the gas hydrates, we first use the dielectric functions shown in Fig.~\ref{diel_CrystallineCO2} to perform calculations for the free energy for an ice film growing on an interface between crystalline CO$_2$ and ice cold water. We see in Fig.~\ref{FreeziingCrystalCO2} that this three layer system has an energy minimum corresponding to an equilibrium ice film with thickness ($d$)  between {3.3\,nm\,\,and\,\,3.9\,nm}, depending on the model for the dielectric function of ice. In the remainder of this letter, we use ice$_\mathrm{JD}$, the Daniels~\cite{Ref12} model for ice,  since both models give very similar results. The thicknesses correlate with the frequency where the dielectric functions of ice and water have a crossing~\cite{MBPhysRevB02017}.

\begin{figure}
\centering
 \includegraphics[width=0.5\columnwidth]{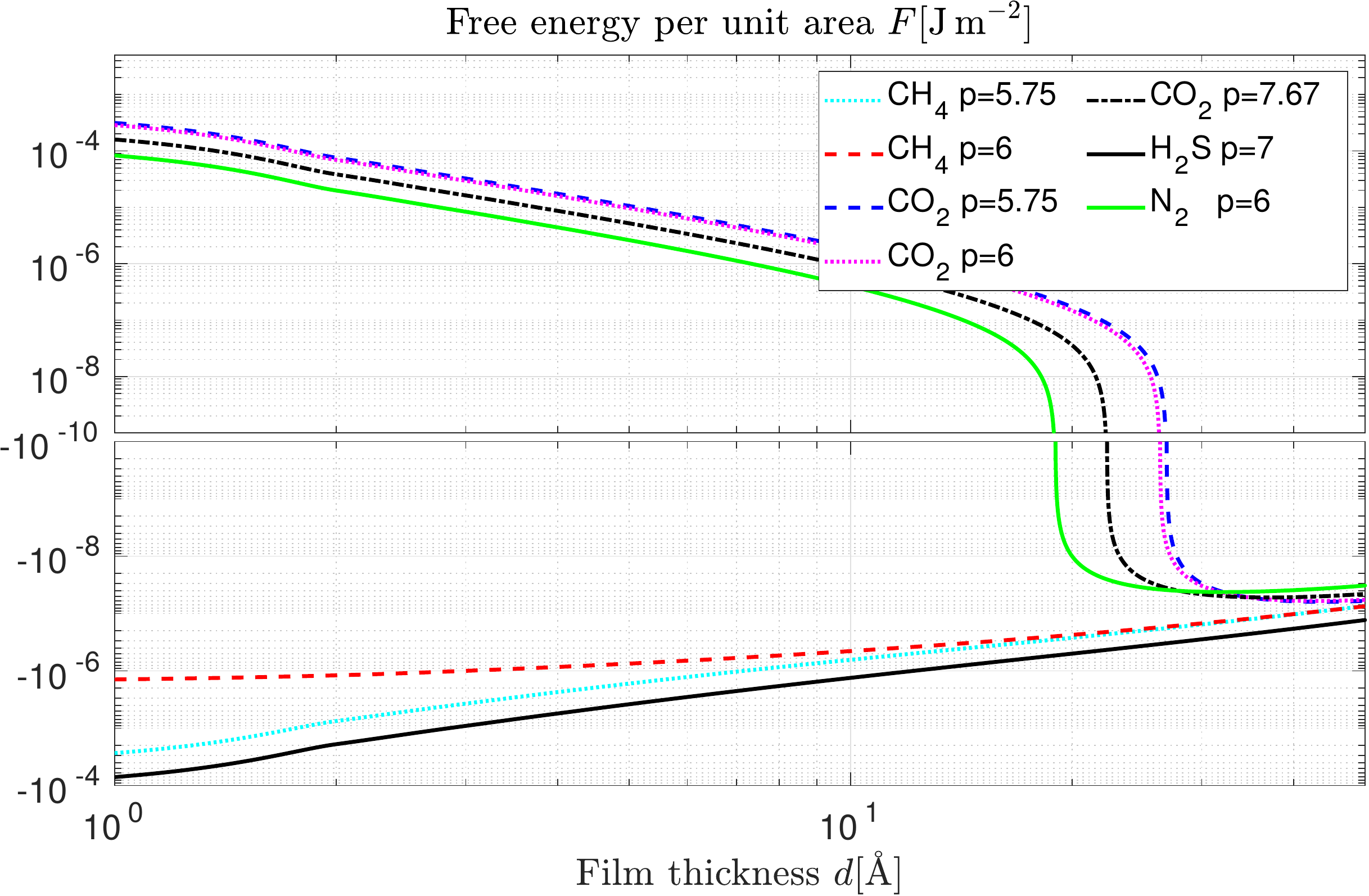}
 
\caption{(Color online) The Lifshitz free energy per unit area (at 273.16 K) for a flat three-layer system (water-ice-gas hydrate) as a function of the ice film thickness $d$. For CO$_2$ gas hydrates and the N$_2$ gas hydrate, energy minimum exist corresponding in each case to an equilibrium ice film thickness: $d_{\mathrm{CO}_2}=$44, 43 and 37\,\AA\ for CO$_2$ volume fractions $p=$ 5.75, 6 and 7.67, respectively, and $d_{\mathrm{N}_2}$=32 \AA.}
 \label{potentials}
\end{figure}
Figure~\ref{potentials} shows the free energy as a function of ice film thickness for different gas hydrates in ice cold water. Ice films are predicted for CO$_2$ hydrates ($d$=44, 43 and 37\,\AA\ for volume fractions $p=$ 5.75, 6 and 7.67, respectively) and for the N$_2$ hydrate ($d=32$\,\AA) but not for any of the CH$_4$ or H$_2$S hydrates. 
In the former cases, retardation plays a role at the nanometer scale as it is the reason for the change in the sign of the Lifshitz energy.
This model is sensitive to the various dielectric functions which are involved in the system~\cite{Zwol2}. While the results are  model dependent for the specific combination of materials used, the clear trend is that interfacial ice caps can exist at some gas hydrates in ice cold water, 
but not for others.

\begin{figure}
\centering
 \includegraphics[width=0.5\columnwidth]{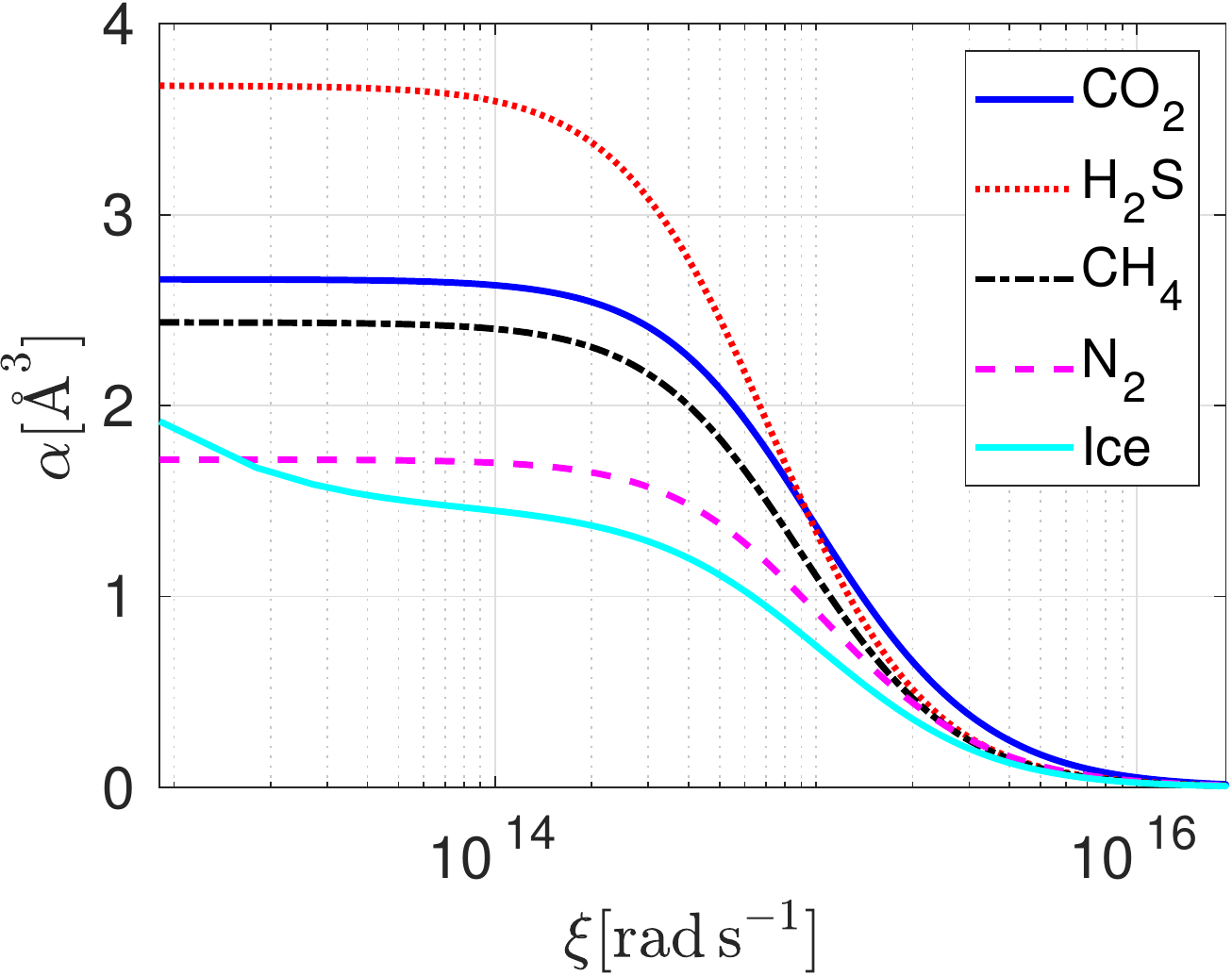}
 \caption{(Color online) Polarisabilities of gas molecules at imaginary frequencies. The polarisability of a water molecule in ice is also shown for comparison. }
 \label{polarisability}
\end{figure}

The stark difference in behaviour between CO$_2$ or N$_2$ hydrates and  CH$_4$ or H$_2$S hydrates can be understood from differences in gas polarisability in the optical/UV  spectrum, combined with the difference in the dielectric spectra of water and ice. These are expressed in the maximum and minimum seen in the product of reflection coefficients in Figure~\ref{reflections}. The positive value  of the product $r_{12} r_{32}$ at low frequencies contributes to stabilisation of the water interface towards the hydrate interface, i.e., wetting, with removal of the ice layer.  Negative values at high frequencies destabilise the water interface, i.e., stabilise the ice layer. The overall behaviour is a balance between these two regimes. As discussed above, the positive and negative regimes  ultimately derive from the reflection coefficient $r_{12}$ between liquid water and ice, that is from the crossing in the dielectric functions of ice and cold water at $1.6 \times 10^{16}\,\rm{ rad\,s}^{-1}$ seen in Figure~\ref{diel_CrystallineCO2}.  The effect of the hydrate (via reflection coefficient $r_{32}$) is to strengthen or attenuate $r_{12}$. Figure~\ref{reflections} shows that the high frequency stabilisation of the ice layer is weaker for CH$_4$ and N$_2$ than for  CO$_2$ at all hydrate ratios, while H$_2$S is only weaker than CO$_2$ at higher water/gas ratios. At low frequencies, destabilisation of the ice layer is much stronger for H$_2$S than CO$_2$, while weaker for N$_2$. In the balance between low frequency destabilisation and high frequency stabilisation of the ice layer, high frequencies dominate for CO$_2$, but are insufficiently weak for CH$_4$. In the case of N$_2$, low frequency behaviour is weaker than for other gases, so again high frequency stabilisation of ice dominates. In the case of H$_2$S, low frequency destabilisation of the ice layer is stronger than for CO$_2$ and dominates over high frequency stabilisation.   These patterns follow the underlying polarisabilities of the gas molecules, see Figure~\ref{polarisability}: the polarisability of CH$_4$ is weaker than CO$_2$ at all frequencies. The polarisability of H$_2$S is significantly stronger than CO$_2$  at low frequencies, but drops rapidly at high frequencies, crossing CO$_2$ to respond similarly to CH$_4$ in the UV spectrum.  The polarisability of N$_2$ is much weaker than other gas molecules, in particular, is much closer to the polarisability of a water molecule. The  polarisability per ice molecule is shown in Figure~\ref{polarisability} for comparison. This results in an N$_2$ gas hydrate dielectric function closer to that of ice, leading to a smaller reflection coefficient. Stabilisation  of the ice layer at a hydrate surface is determined predominantly from the  polarisability of the gas molecule relative to a water molecule in the optical spectrum around $3\times 10^{15} \,\rm{rad\, s}^{-1}$ (stabilising water wetting) and in the UV spectrum around $3\times 10^{16}\,\rm{ rad\,s}^{-1}$ (stabilising the ice layer). 
 
\subsection{Size dependence for floating of gas hydrate clusters}

Buoyancy of gas hydrate particles is of considerable importance for understanding the distribution and composition of ices, water and gases in subglacial water bodies in Antarctica and on ocean bearing moons of our solar systems and extra solar planets. Buoyancy of gas hydrates on these water bodies depends on hydrate density and assumed ocean densities.

Lake Vostok, located 4\,km below the Antarctic surface, is an analogue of deeper subglacial oceans the Jovian and Saturnian moons and is a notable target for astrobiological studies. McKay et al.\,\cite{McKay2003} suggest the observed lack of gas hydrates accreted at the top of Lake Vostok in Antarctica (density 1.016\,g\,cm$^{-3}$) is consistent with formation of relatively dense CO$_2$ clathrate hydrates that sink to the lake floor. Mousis et al. (2013) later estimated the densities of type I and II clathrate CO$_2$ hydrates in Lake Vostok, concluding that CO$_2$-containing type I clathrates sink above a critical CO$_2$ composition in the lake.

\begin{figure}
\centering
 \includegraphics[width=0.5\columnwidth]{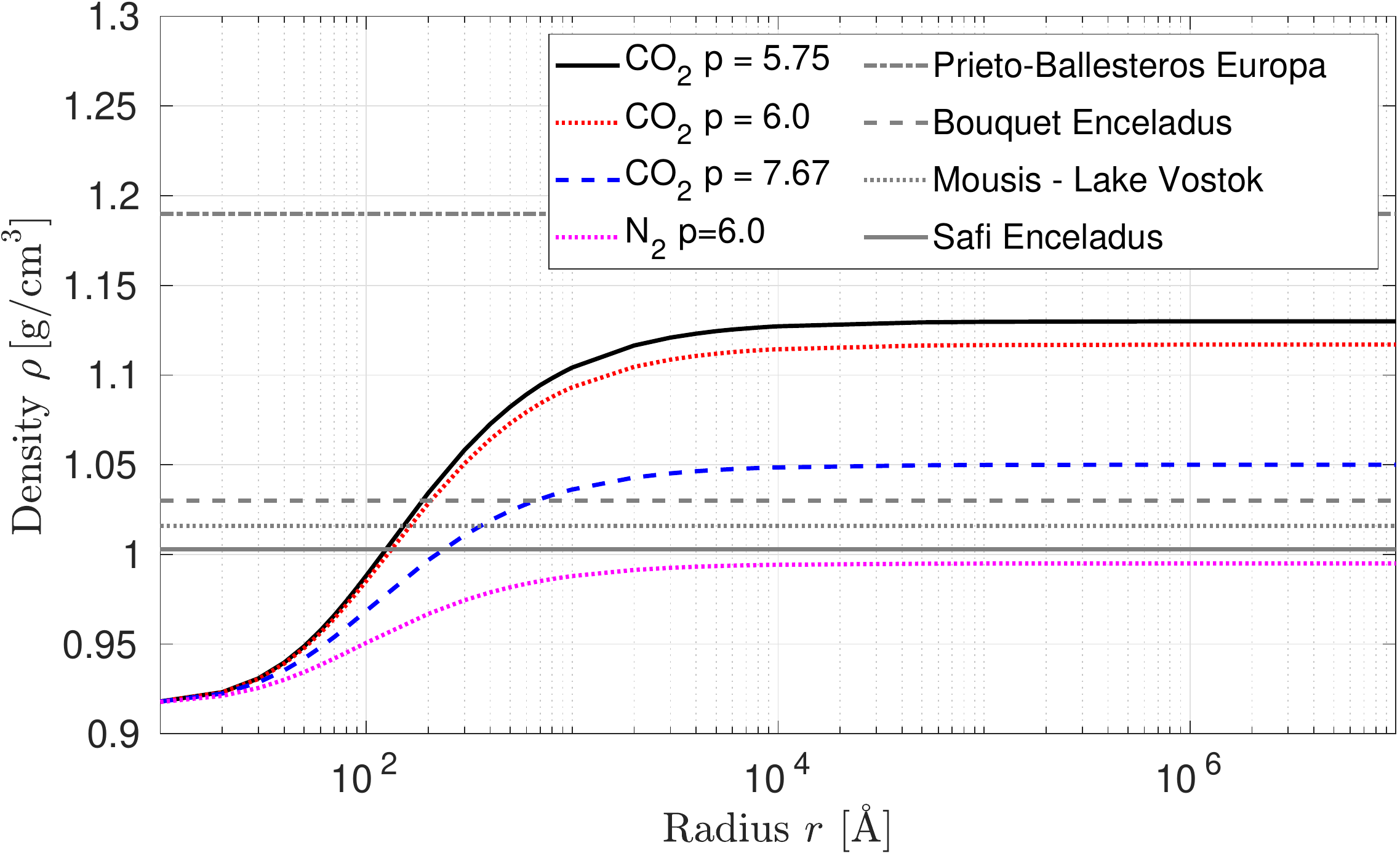}
 \caption{(Color online) The average densities of gas hydrate particles of varying radius for CO$_2$ hydrates with density fraction $p=5.75$ (solid black line), $p=6.0$ (dotted red line) and $p=7.67$ (dashed blue line) and N$_2$ with $p=6.0$ (dotted magenta line), each compositional variant coated in a layer of ice of thickness determined by the polarizability and amount of the entrapped gas species as specified above. As comparison we show the water densities in different systems. References for the water/gas number density ratios $p$ are given in Tab.~\ref{EduTable}. }
 \label{Avdensity}
\end{figure}

Prieto-Ballesteros et al.\,\cite{PrietoBallesteros2005} considered buoyancy of type I CO$_2$, SO$_2$, CH$_4$ and H$_2$S gas hydrates (space group 223) on Europa, where two extreme models for the density of the ocean water were considered, namely a eutectic brine of composition MgSO$_4$-H$_2$O system with density 1.19\,g\,cm$^{-3}$ and a low salinity water ocean of density 1.0\,g\,cm$^{-3}$. Safi et al.~\cite{Safi2017} discuss the buoyancy of CO$_2$ hydrates using two further density estimates of the oceans of Europa (density  1.016\,g\,cm$^{-3}$) and Enceladus (1.003\,g\,cm$^{-3}$) and used measured gas hydrate densities. Bouquet et al.~\cite{Bouquet2015} consider the buoyancy of multiple guest clathrates on Enceladus using a calculated high pressure sea water density of 1.030\,g\,cm$^{-3}$, concluding the type I clathrates are marginally denser (1.040\,g\,cm$^{-3}$) than the sea water and type II significantly lighter (0.970\,g\,cm$^{-3}$). 

Figure~\ref{Avdensity} shows the average densities of gas hydrate particles of varying radius, each compositional variant coated in a layer of ice of thickness determined by the polarizability and amount of the entrapped gas species. These were calculated using equation ~\ref{Equrho} below. The equilibrium ice film thicknesses and densities are given in the text immediately below Figure~\ref{potentials}. The value used for the density of pure ice was 0.9167\,g\,cm$^{-3}$. Horizontal lines represent the estimated or measured density of the ocean/sea water on the various bodies.

It is quite apparent that for CO$_2$ gas hydrate particles, which are otherwise denser than most models for the water on Enceladus or in Lake Vostok, an equilibrium ice layer of the order of several nm has a significant impact on the buoyancy of the particles. When the radii of these is below approximately 20-100\,nm, the average density drops below values estimated for the ocean water density on Enceladus and consequently will float. We use the following simple expression for the average density of a ice coated gas hydrate cluster (approximated as a sphere),
\begin{equation}
\rho_{\rm av}={\frac{\rho_{\rm h} r_{\rm h}^3+\rho_{\rm i}  [(r_{\rm h}+d)^3-r_{\rm h}^3 ]}{(r_{\rm h}+d)^3}}\,. 
\label{Equrho}
\end{equation}
Here $\rho_{\rm av}$ is the average density of mixed particle comprising a clathrate hydrate core and a shell of water ice, $\rho_{\rm h}$ densities are given in Tab.~\ref{EduTable} for different gas hydrates, $\rho_{\rm i}$ is given for ice above, and $r_{\rm h}$ the radius of a gas hydrate cluster. Finally,
$d$ is the approximate thickness of each ice film at planar water-CO$_2$ gas hydrate and water-N$_2$ gas hydrate interfaces given above.

\section{Conclusions}

In analogy to the premelting layers of ice~\cite{Elbaum,Wettlaufer,Dash1995}, we found that freezing of gas hydrates in  ice cold water is caused by an energy minimum in dispersion energies. This is not expected at water surfaces~\cite{Elbaum2} but predicted to occur at some water-solid interfaces~\cite{MBPhysRevB02017}. 
We find that a significant difference between different gas hydrate surfaces in water lies in whether they are coated with a nano-sized interfacial ice cap or not. The result is sensitive to the details in the dielectric functions of the materials involved. However, our results indicate that some hydrates are more likely to have interfaces that are kept dry by an interfacial ice cap. We have seen this trend for three different volume fractions of CO$_2$ hydrates in water as well as for N$_2$ hydrate and crystalline CO$_2$ in water. Other hydrates, CH$_4$ hydrate in water, are more likely to stay wet and have no interfacial ice cap. 

A review~\cite{Maeda} a few years ago asked the question if gas hydrate surfaces in  air are dry or wet. Our results are consistent with  gas hydrate surfaces that are in equilibrium with water molecules in vapor phase. If  a film of water is adsorbed on a gas hydrate surface, much thicker than say 10\,nm, then our calculations can be extended to predict that a fraction of that water will form an interfacial ice layer between the water film and the surface of the CO$_2$ (or N$_2$) hydrates but not so for CH$_4$ (or H$_2$S) hydrates. 
These differences for materials, whether their interfaces stay dry or wet, are expected to influence the fluxes of gas molecules into the liquid water and then further towards the surrounding atmosphere.
Further, as we discussed above a dry surface may affect the growth and overall density of gas hydrate crystals. The density of type I CO$_2$ hydrate crystal densities are similar to that predicted for different ice coated ocean waters on Earth, Enceladus and Europa~\cite{McKay2003,PrietoBallesteros2005,Mousis2013,Safi2017,Bouquet2015}.
The density values in Tab.~\ref{EduTable}  for CO$_2$ hydrate suggest a water ice cap layer could make a significant difference in buoyancy when hydrate crystals have diameters in the range of approximately {20-100\,nm}, based on the {3-4\,nm} ice films we predict. Indeed if a layer of interfacial ice cap grows on a hydrate crystal early after nucleation, its growth may be restricted to such small sizes, leading to the formation of nanoscale, ice-capped CO$_2$ hydrate crystals with positive buoyancy.

Besides the requirement of accurate dielectric functions for quantitative predictions of such ice layer thicknesses, the restriction to interactions caused by dispersion forces yields a source of uncertainties. For non-polar systems, it would be sufficient to neglect electrostatic effects. However, water is a polar medium, thus interactions caused by permanent dipole moments will also play a role and will shift this theory to a more precise one. The extension of the theory of dispersion forces to include permanent dipole moments is of current interest for several groups and will also be part of further investigations. However, a simple estimation of such effects shows a small contribution to the dispersion forces which is smaller than in the vacuum case due to the shielding effect of the environmental medium.

We have notably shown that the above-mentioned density dependence of the gas hydrates induce a sinking or floating of the particles which is important for carbon capture and storage via gas hydrates~\cite{doi:10.1093/gji/ggv015}. The creation
of an interfacial ice layer modify the average density of
the particle, thus the buoyancy that determines the floating or sinking of the particle.
When studying very small gas hydrates, the particle’s curvature may be expected to play a role.
It can easily be incorporated into theory by changing the geometry from a planar to a spherically
layered system. However, since the size of the gas hydrate clusters are much larger than the predicted ice film layer a planar approximation is expected to give useful estimates. Such investigations will also effect the description of crystallization processes in particular for cloud creation~\cite{doi:10.1175/2007JAS2515.1} by treating the gas hydrate as cloud condensation nuclei.

\section*{Acknowledgement}

We gratefully acknowledge support from the Research Council of Norway (Project  250346), the German Research Council (grant BU 1803/6-1, S.Y.B. and J.F., BU 1803/3-1, S.Y.B.), the Research Innovation Fund by the University of Freiburg (S.Y.B., J.F.),  the Freiburg Institute for Advanced Studies (S.Y.B.), and FAPERJ (JCNE E-26/203.223/2016).
DFP acknowledges the grant of resources from the National Computational
Infrastructure (NCI), which is supported by the Australian
Government. Data available from authors.
\section*{Disclaimer}
This document is the unedited Author’s version of a Submitted Work that was subsequently accepted for publication in ACS Earth and Space Chemistry, copyright © American Chemical Society after peer review. To access the final edited and published work see:

\url{https://doi.org/10.1021/acsearthspacechem.9b00019} 

\bibliographystyle{unsrt}
\bibliography{gasbib}

\end{document}